\documentclass[nofootinbib,aps,amssymb,floatfix,superscriptaddress,preprintnumbers]{revtex4}

\usepackage{graphicx}
\usepackage{bm}
\usepackage{amsmath}
\usepackage{amssymb}
\usepackage{amsfonts}
\usepackage{float}
\usepackage{hyperref}
\usepackage{dsfont}  
\usepackage{slashed}  
\usepackage{booktabs}
\usepackage{multirow}
\usepackage{subfigure}
\usepackage[sort&compress]{natbib}
\usepackage{xcolor}
\usepackage{ulem}
\usepackage[percent]{overpic}

\newcommand{\be}{\begin{equation}}  
\newcommand{\ee}{\end{equation}}  
\newcommand{\beq}{\begin{eqnarray}} 
\newcommand{\eeq}{\end{eqnarray}}

\newcommand{\Slash}[1]{{\ooalign{\hfil/\hfil\crcr$#1$}}}
\newcommand{\nn}{\nonumber \\}

\newcounter{RSQ}

\begin{document}

\title{ Mass radius and D-term of atomic nuclei in relativistic mean field theory}

\author{Yoshitaka Hatta  }
\affiliation{Physics Department, Brookhaven National Laboratory, Upton, NY 11973, USA}
\affiliation{RIKEN BNL Research Center, Brookhaven National Laboratory, Upton, NY 11973, USA}
%\email{yhatta@bnl.gov}

\author{Tomohiro Oishi}

\affiliation{Ibaraki College in National Institute of Technology
(KOSEN), Hitachinaka 312-8508, Japan}

\author{Makoto Oka}
\affiliation{Nishina  Center for Accelerator-Based Science, RIKEN, Wako 351-0198, Japan}
\affiliation{Department of Physics, Tohoku University, Sendai 980-8578, Japan}
\affiliation{Advanced Science Research Center, Japan Atomic Energy Agency, Tokai, Ibaraki, 319-1195, Japan}
\date{\today}

\begin{abstract}
Based on  relativistic mean field theory for atomic nuclei, we compute the mass radius and other radii associated with the energy momentum tensor for dozens of spin-0 nuclei across the nuclear chart. We also compute the D-term of these nuclei, the forward limit of the gravitational form factor $D(t=0)=D$. The dependence  on  the  neutron number $N$ is systematically studied for calcium (Ca), nickel (Ni), zirconium (Zr), tin (Sn) and lead (Pb) isotopes. Remarkably, $|D|$ does not monotonically increase  with $N$. Instead, it  exhibits local maxima and minima   when  $N$  equals a magic number and even a sub-magic number. 
This results in characteristic kinks in the mass, scalar, tensor and shear radii of these isotopes. 
Our work for the first time elucidates the strong sensitivity of the various mechanical properties of nuclei to the nuclear shell structure.

\end{abstract}

\maketitle

\section{Introduction}

The history of measuring the size of atomic nuclei is nearly as old as nuclear physics itself. Already in 1911, Rutherford  concluded that the size of the gold nucleus is about $10^4$-$10^5$ times smaller than that of the gold atom. In the 1930s, the liquid drop model led to the well-known empirical formula for the nuclear radius $r$ as a function of the atomic number $A$ 
\beq
r=r_0A^{\frac{1}{3}}, \qquad r_0\approx 1.2\ {\rm fm}
\eeq
still widely used in textbooks. This formula reflects the saturation property of nuclear matter, namely the volume is proportional to the number of nucleons $V\propto r^3\propto A$, as a consequence of the short-ranged nature of the nuclear force. 
With the advent of electron scattering in the 1950s and the subsequent experimental efforts over several decades, the measurement of nuclear radii has become precision science. For many nuclei, the root-mean-square  charge radius $r_{em}\equiv \sqrt{\langle r^2\rangle_{em}}$ has been determined with percent-level accuracy or even better  \cite{DeVries:1987atn,Angeli:2013epw}. For example, the most recent measurement of the charge radius of the lead nucleus $\,^{208}$Pb reported $r_{em}=5.5062(17)$ fm, with less than 0.1\% errors \cite{Sun:2025qll}. 

The charge radius is defined through the electromagnetic form factor $\langle p'|j_{em}^\mu |p\rangle$ where $j^\mu_{em}=\sum_f e_f \bar{q}_f\gamma^\mu q_f$ is the electromagnetic current operator in QCD. In low energy nuclear physics, this may be roughly approximated by  the proton current $j_{em}^\mu \approx \bar{P}\gamma^\mu P$. As such, the charge radius predominantly  reflects the distribution of protons, even though neutrons are increasingly more numerous in larger nuclei. This motivates one to consider more inclusive measures of  nuclear size that  do not  distinguish protons and neutrons. For example, the half-density radius $R$ of  the Woods-Saxon density profile $\rho(r)\sim  \left(1+\exp{(r-R)/a}\right)^{-1}$   \cite{Woods:1954zz} may be regarded as such a radius, although its definition and extraction are model-dependent. More fundamentally, one can define a radius  rigorously in QCD through   the form factor of the baryon number current $\langle p'|j_B^\mu|p\rangle$, or the gravitational form factor  (GFF)  $\langle p'|T^{\mu\nu}|p\rangle$ \cite{Kobzarev:1962wt,Pagels:1966zza,Polyakov:2018zvc}, where $T^{\mu\nu}$ is the energy momentum tensor (EMT).   
Conserved operators such as  $j_B^\mu$ and $T^{\mu\nu}$ have clear physical interpretations and can be straightforwardly constructed not only in QCD, but also in any effective models of quarks, nucleons and mesons with a relativistic Lagrangian formulation. The first calculation of the `mass radius' $r_m^2\sim \int d^3r r^2 T^{00}$ and the `scalar radius'  $r_s^2\sim \int d^3r r^2 T^\mu_\mu$  was done in this way   for the proton \cite{Goeke:2007fp,Cebulla:2007ei}. There are a number of more recent calculations for hadrons   \cite{Jung:2013bya,Burkert:2018bqq,Hatta:2018ina,Lorce:2018egm,Anikin:2019kwi,Kharzeev:2021qkd,Mamo:2022eui,Fujita:2022jus,Won:2022cyy,Wang:2022uch,Duran:2022xag,Guo:2023pqw,Hackett:2023rif,Wang:2024lrm,Cao:2024zlf,Goharipour:2025yxm,Broniowski:2025ctl,Fujii:2025aip,Tanaka:2025pny,Stegeman:2025sca,Sugimoto:2025btn,Nair:2025sfr,Sain:2025kup,Hippelainen:2026izm,Fukushima:2026wwc} and light nuclei such as the deuteron and the helium-4 \cite{Freese:2022yur,He:2023ogg,He:2024vzz,Cosyn:2026gyy}.

The study of the energy momentum tensor for larger nuclei requires   more collective methods appropriate for many-body systems  \cite{Polyakov:2002yz,Guzey:2005ba,Liuti:2005qj,GarciaMartin-Caro:2023klo,GarciaMartin-Caro:2023toa}. 
In \cite{GarciaMartin-Caro:2023klo,GarciaMartin-Caro:2023toa}, one of the present authors and  collaborators employed the Skyrme model \cite{Skyrme:1961vq} to compute the GFFs and  associated radii for a number of atomic nuclei.  In this model, a nucleus is mimicked by a `Skyrmion,' a finite-energy configuration of the pion field with a protected topological number identified with the atomic number $A$ \cite{Braaten:1988cc,Carson:1991fu,Manko:2007pr,Feist:2012ps,Gudnason:2022jkn,Hal,Freire:2025mkf}. For small Skyrmions, say  $A\lesssim 16$ \cite{Gudnason:2022jkn,Gudnason:2026wbe}, such configurations have been systematically constructed and classified in order to describe nuclear structure.  However,  numerical construction of Skyrmions becomes increasingly difficult as $A$ grows. Solutions with $A\sim {\cal O}(100)$ do exist \cite{Feist:2012ps,Hal,GarciaMartin-Caro:2023nty,Freire:2025mkf}, but their correspondence to actual heavy nuclei is primarily qualitative without including further corrections.

In this paper, we employ the relativistic  mean field model (RMF) to compute the mass radius and other related radii of spherical, even-even  nuclei. RMF was originally developed for nuclear matter (`Walecka model') \cite{Walecka:1974qa}, and later applied to finite nuclei \cite{Serot:1979cc,Bender:2003jk,Niksic:2011sg}.  It is one of the most successful, time-tested models for medium-sized  or larger nuclei $A\gtrsim 16$ in low energy nuclear physics. In contrast to the Skyrme model, RMF works better for heavier nuclei, and even the region $A\sim 200$ poses no technical difficulties. The model is formulated as a relativistic field theory   in which nucleons are represented by Dirac particles self-consistently moving in the background potential created by mesons. 
In a pioneering work \cite{Guzey:2005ba},  Guzey and Siddikov  employed the original Walecka model  to calculate the so-called D-term \cite{Polyakov:2018zvc}, the forward $(t=0)$ limit of the GFF $D(t)$,  for several nuclei. 
We will use one of the most advanced versions of RMF called \texttt{DIRHB}    \cite{Niksic:2014dra} designed for the ground state of even-even nuclei.   \texttt{DIRHB} is a  package written in Fortran, and is able to reproduce the basic properties of hundreds of spherical and deformed nuclei using ${\cal O}(10)$  parameters.
In particular, the charge radii and binding energies agree with the experimental data within a few percent or even less. One may then expect that the model can also provide   realistic predictions  for quantities associated with the energy momentum tensor  $T^{\mu\nu}$. The result offers novel perspectives on the size and internal structure of atomic nuclei that can be explored in future experiments. 

This paper is organized as follows. 
In section II, we present the RMF Lagrangian, the equation of motion, the energy momentum tensor and related quantities. Definitions of various radii and the D-term are summarized in section III. Results of numerical analyses are presented in section IV. We conclude in section V. Appendix presents a review of the Dirac equation in the spherical coordinate system.

\section{Relativistic mean field theory}

Our starting point is the following Lagrangian describing the nucleons, mesons and the photon  
\beq
{\cal L}&=& \bar{\psi}(i\Slash \partial-m)\psi -\frac{1}{4}\Omega_{\mu\nu}\Omega^{\mu\nu}+\frac{m^2_\omega}{2}\omega_\mu \omega^\mu - g_\omega\omega_\mu\bar{\psi}\gamma^\mu \psi  \nn
&& -\frac{1}{4}\vec{R}_{\mu\nu}\cdot \vec{R}^{\mu\nu}+\frac{m^2_\rho}{2}\vec{\rho}_\mu \cdot\vec{\rho}^\mu - g_\rho\vec{\rho}_\mu \cdot\bar{\psi}\gamma^\mu \vec{\tau}\psi  \nn
&& +\frac{1}{2}\partial_\mu \sigma \partial^\mu \sigma -\frac{m_\sigma^2}{2}\sigma^2-g_\sigma \sigma\bar{\psi}\psi \nn
&& -\frac{1}{4}F^{\mu\nu}F_{\mu\nu} - eA_\mu\bar{\psi}\gamma^\mu\frac{1+\tau^3}{2}\psi. \label{lag}
\eeq
Our metric convention is  $g^{\mu\nu}={\rm diag}(1,-1,-1,-1)$. 
$\Omega_{\mu\nu}=\partial_\mu \omega_\nu - \partial_\nu \omega_\mu$, $\vec{R}_{\mu\nu}=\partial_\mu \vec{\rho}_\nu - \partial_\nu \vec{\rho}_\mu$ and $F_{\mu\nu}=\partial_\mu A_\nu -\partial_\nu A_\mu$ represent the $\omega$-meson, the $\rho$-meson and the electromagnetic field, respectively. $\sigma$ is the $\sigma$-meson field. $\tau^{i=1,2,3}$ are  the Pauli matrices. $\psi$ is the $A=N+Z$ component nucleon Dirac spinor  with $\tau^3=1$ for  protons and $\tau^3=-1$ for  neutrons.  $m=939$ MeV is the nucleon mass. The previous work based on the Walecka model \cite{Guzey:2005ba} included the $\omega$ and $\sigma$ mesons. Here and in the rest of this section, we also consider the $\rho$ meson and the photon which are included by default in the DIRHB code. However, we will neglect the photon field in actual numerical calculations for a reason to be explained later.

RMF in general assumes independent motion of nucleons in self-consistent mean fields,  where each nucleon occupies a single particle state (labeled by $a$) with the single particle energy $\epsilon_a$ (including the nucleon mass) and the Dirac wave function, $\psi_a(x_a)$.
The Lagrangian (\ref{lag}) allows the following conserved currents, the baryonic current $j_B^{\mu}=\bar\psi\gamma^{\mu} \psi$, and the isospin current, $\vec j_I^{\mu} =\bar\psi\gamma^{\mu}\vec\tau\psi$.% 
\footnote{The isospin current is not fully conserved due to  the Coulomb interaction, while the $\tau_3$ part, $Z-N$, is conserved.}
The electromagnetic current is also conserved, $j_{em}^{\mu} = (1/2) (j_B^{\mu}+j_{I3}^{\mu})$, where the proton is treated as a point charge. 
We define the corresponding  vector ($B$) and isovector-vector ($I$) charge densities as 
\beq
&&\rho_{B}  =\langle A| \bar{\psi}\gamma^0\psi|A\rangle=\sum_{a=1}^{A} \psi^{\dagger}_{a} \psi_{a},
\nn
&& \rho_{I} %=\sum_{a=1}^{A} \psi^{\dagger}_{a} \tau^{3} \psi_{a} 
=\langle A| \bar{\psi}\gamma^0\tau^3\psi|A\rangle=\sum_{a \in Z} \psi^{\dagger}_{a} \psi_{a} -\sum_{a \in N} \psi^{\dagger}_{a} \psi_{a}. \nn
&&\rho_{s}  %=\sum_{a=1}^{A} \bar{\psi}_{a} \psi_{a} 
=\langle A| \bar{\psi}\psi|A\rangle =\sum_{a=1}^{A} \psi^{\dagger}_{a} \gamma_{0} \psi_{a},
\eeq
where $|A\rangle=|A(Z,N)\rangle$ denotes the ground state of the nucleus of the mass number $A=Z+N$.
We have also introduced the scalar ($s$) density $\rho_s$, which is not associated with a conserved current.

The present DIRHB model introduces density dependent coupling constants  \cite{toki} in order to reproduce the binding energies of  nuclei in a wide range of the nuclear chart. 
Namely, the meson-nucleon coupling constants are chosen to depend on the baryon density $g_\sigma= g_{\sigma}(\rho_{B})$, $ g_{\omega}=g_{\omega}(\rho_{B})$ and $g_\rho= g_{\rho}(\rho_{B})$ (see (\ref{grho}) below).  
The equation of motion for the nucleon field is obtained by varying the Lagrangian 
\beq
i\Slash \partial \psi -(m+g_\sigma \sigma)\psi-\left(g_\omega\omega_\mu  +g_\rho \vec{\rho}_\mu\cdot \vec{\tau} +\Sigma^R_\mu +eA_\mu \frac{1+\tau^3}{2}\right)\gamma^\mu \psi     =0,
\label{eom}
\eeq
where $\Sigma^R_\mu$ is the so-called rearrangement contribution  \cite{Niksic:2014dra}
\beq
\Sigma^R_\mu = \frac{j_{B\mu}}{\rho_B} \left[ \sigma  \frac{\partial g_{\sigma}}{\partial \rho_B} \rho_{s}
+\omega^{\nu} \frac{\partial g_{\omega}}{\partial \rho_B} j_{B\nu}
+\vec{\rho}^{\nu}  \frac{\partial g_{\rho}}{\partial \rho_B} \vec{j}_{I \nu}  \right].
\eeq
%with $j_\mu = \bar{\psi}\gamma_\mu \psi$. 
The rearrangement term typically shifts all single particle levels by tens of MeV.

In the following, we will only consider static, spherically symmetric solutions having in mind the ground state of  spin-0 nuclei. 
 Each nucleon wave function has time dependence 
\beq
\psi_a(t,\vec{r})= e^{-i\epsilon_at}\psi_a(\vec{r}),
\eeq
according to the single-particle energy (including the nucleon mass). 
In contrast, the meson and electromagnetic fields
depend only on the spatial coordinate. 
Furthermore, the rotational symmetry permits only the time component of the vector
fields being non-zero and spherically symmetric, $\omega_{\mu=0}\equiv \omega(r)$, $\vec\rho_{\mu=0}\equiv (0,0,\rho(r))$, and $A_{\mu=0}=\phi(r)$ (Coulomb field).
They  satisfy the equations $(r=|\vec{r}|$) 
\beq
&&(\vec{\nabla}^2-m_\sigma^2)\sigma(r) = g_\sigma \rho_s(r),%\bar{\psi}\psi 
\nn 
&&(\vec{\nabla}^2-m_\rho^2)\rho(r) =- g_\rho \rho_{I}(r),%\bar{\psi}\gamma^0\tau^3\psi
\nn
&&(\vec{\nabla}^2-m_\omega^2)\omega(r) =- g_\omega \rho_B(r),
%\bar{\psi}\gamma^0\psi 
\nn 
&&\vec{\nabla}^2\phi(r)=-\frac{e}{2}(\rho_B(r)+\rho_{I}(r)).
%\bar{\psi}\gamma^0\frac{1+\tau^3}{2}\psi 
\label{meson}
\eeq
Similarly, only the time component of the rearrangement term survives 
\beq
\Sigma_{\mu=0}^{R} =\sigma \frac{\partial g_{\sigma}}{\partial \rho_B} \rho_{s} +\omega  \frac{\partial g_{\omega}}{\partial \rho_B} \rho_{B} + \rho \frac{\partial g_{\rho}}{\partial \rho_B} \rho_{I} . \label{Sigma0}
\eeq

The symmetric (Belinfante-improved) energy momentum tensor is most conveniently calculated by embedding the action $S=\int d^4x \sqrt{-g} {\cal L}$ in a curved spacetime and differentiating 
\beq
T_{\mu\nu}= \frac{2}{\sqrt{-g}}\frac{\delta S}{\delta g^{\mu\nu}}.
\eeq
The result is 
\beq
T_{\mu\nu} &=& 
 \bar{\psi}i\gamma_{(\mu} \overleftrightarrow{\partial}_{\nu)} \psi- g_{\mu\nu} \bar{\psi}\Sigma^R_\alpha \gamma^\alpha\psi  \nn 
&&
-\Omega_{\mu\rho}\Omega_\nu^{\ \rho} +m_\omega^2\omega_\mu \omega_\nu -g_\omega\omega_{(\mu}\bar{\psi}\gamma_{\nu)} \psi  -g_{\mu\nu}\left(-\frac{1}{4}\Omega_{\alpha\beta}\Omega^{\alpha\beta}+\frac{m^2_\omega}{2}\omega_\alpha \omega^\alpha \right) \nn 
&& -\vec{R}_{\mu\rho}\cdot \vec{R}_\nu^{\ \rho} +m_\rho^2\vec{\rho}_\mu\cdot \vec{\rho}_\nu -g_\rho\vec{\rho}_{(\mu}\cdot \bar{\psi}\gamma_{\nu)}\vec{\tau} \psi  -g_{\mu\nu}\left(-\frac{1}{4}\vec{R}_{\alpha\beta}\cdot \vec{R}^{\alpha\beta}+\frac{m^2_\rho}{2}\vec{\rho}_\alpha \cdot \vec{\rho}^\alpha\right) \nn 
&& +\partial_\mu \sigma \partial_\nu \sigma -g_{\mu\nu}\left( \frac{1}{2}\partial_\alpha \sigma \partial^\alpha \sigma -\frac{m_\sigma^2}{2}\sigma^2 \right)  -F_{\mu\rho}F_\nu^{\ \rho}-eA_{(\mu}\bar{\psi}\gamma_{\nu)}\frac{1+\tau^3}{2}\psi +\frac{g_{\mu\nu}}{4}F^{\alpha\beta}F_{\alpha\beta} ,    \label{tm}
\eeq
where $A_{(\mu}B_{\nu)}=\frac{A_\mu B_\nu+A_\nu B_\mu}{2}$ and $\overleftrightarrow{\partial}_\mu = \frac{1}{2}(\overrightarrow{\partial}_\mu - \overleftarrow{\partial}_\mu)$, and we have already imposed the Dirac equation. 
The energy density reads 
\beq  
T^{00} &=& \bar{\psi}i\gamma^0\overleftrightarrow{\partial}_0\psi +\frac{1}{2}\partial_k \omega \partial_k \omega + \frac{m_\omega^2}{2} \omega^2 -g_\omega\omega \bar{\psi}\gamma^0\psi 
+ \frac{1}{2}\partial_k \sigma \partial_k \sigma + \frac{m_\sigma^2}{2}\sigma^2\nn && +\frac{1}{2}\partial_k\rho \partial_k \rho+\frac{m_\rho^2}{2}\rho^2-g_\rho \rho \bar{\psi}\gamma^0\tau^3\psi   +\frac{1}{2}\partial_k \phi \partial_k \phi-e\phi\bar{\psi}\gamma^0\frac{1+\tau^3}{2} \phi -  \Sigma^R_0\bar{\psi} \gamma^0\psi \nn
&=& \sum_{a=1}^A\epsilon_a \psi^{\dagger}_a (r) \psi_a(r)
+ \left[ \frac{1}{2}\left(\frac{d\omega}{dr}\right)^2  + \frac{m_\omega^2}{2} \omega^2 -g_\omega \omega\bar{\psi}\gamma^0\psi  \right]
+\left[ \frac{1}{2}\left(\frac{d\sigma}{dr}\right)^2+ \frac{m_\sigma^2}{2}\sigma^2  \right]  \nn
&& +\left[ \frac{1}{2}\left(\frac{d\rho}{dr}\right)^2+\frac{m_\rho^2}{2}\rho^2-g_\rho \rho\bar{\psi}\gamma^0\tau^3\psi \right]
+\left[ \frac{1}{2} \left(\frac{d\phi}{dr}\right)^2 -e\phi\bar{\psi}\gamma^0\frac{1+\tau^3}{2}\psi \right]
-  \Sigma^R_0\bar{\psi} \gamma^0\psi,   \label{t00}
\eeq
where in the second equality we assumed spherical symmetry.  
$T^{00}$ differs from the energy density ${\cal H}$ normally used in DIRHB, see Eq.~(8) of \cite{Niksic:2014dra}. The latter is the $00$-component of the canonical energy momentum tensor ${\cal H}=T^{00}_{can}$ with 
\beq
T^{can}_{\mu\nu}= i\bar{\psi}\gamma_\mu\partial_\nu\psi +\partial_\mu \sigma \partial_\nu \sigma-\Omega_{\mu\alpha} \partial_\nu \omega^\alpha -R_{\mu\alpha}\partial_\nu \rho^\alpha -F_{\mu\alpha}\partial_\nu A^\alpha -g_{\mu\nu}{\cal L} ,
\eeq
derived from the Lagrangian (\ref{lag}) via the usual Noether method. One can check that the difference reduces to  total derivative terms so that their spatial integrals are equal 
\beq
M=Am+{\cal E}=\int d^3r T^{00}(r) = \int d^3r T^{00}_{can}(r), \label{M}
\eeq
where $M$ is the total mass of the nucleus, i.e. $A$ times the nucleon mass $m$ plus the nuclear binding energy ${\cal E}(<0)$.

In addition to the energy density, let us also consider  the trace of the energy momentum tensor 
\beq
T^\mu_\mu &=& 
%T^{00}-T_{ii} \nn  &=&
\bar{\psi}i\overleftrightarrow{\Slash \partial}\psi -m_\omega^2\omega^2-g_\omega\omega \bar{\psi}\gamma^0\psi  -\partial^\mu \sigma \partial_\mu \sigma +2m_\sigma^2\sigma^2 -g_\rho \rho\bar{\psi}\gamma^0\tau^3\psi -m_\rho^2\rho^2  -e\phi\bar{\psi}\gamma^0\frac{1+\tau^3}{2}\psi - 4\Sigma^R_0\bar{\psi} \gamma^0\psi  \nn
&=& m\bar{\psi}\psi   -m_\omega^2\omega^2  +\left(\frac{d \sigma}{dr}\right)^2 +2m_\sigma^2\sigma^2 +g_\sigma \sigma\bar{\psi}\psi  -m^2_\rho \rho^2   -3\Sigma^R_0\bar{\psi} \gamma^0\psi,  \label{trace}
\eeq
where in the second equality we used the Dirac equation. In QCD, the trace is nonzero due to the QCD trace anomaly. In the present model, it is effectively expressed by the nucleon and meson fields. 
Moreover, from the spatial component 
\beq
T_{ij} &=& \frac{i}{4}\bar{\psi}\left(\gamma_{i} \partial_j+\gamma_j\partial_i -\gamma_i\overleftarrow{\partial}_j -\gamma_j\overleftarrow{\partial}_i\right) \psi 
-\partial_i\omega \partial_j\omega+\frac{\delta_{ij}}{2}\left(\partial_k \omega \partial_k\omega+m^2_\omega\omega^2\right)  \nn 
&& +\partial_i \sigma \partial_j \sigma +\frac{\delta_{ij}}{2}\left( -\partial_k \sigma \partial_k \sigma -m_\sigma^2\sigma^2\right) -\partial_i\rho \partial_j\rho+\frac{\delta_{ij}}{2}\left(\partial_k \rho \partial_k\rho+m^2_\rho\rho^2\right) \nn 
&& -\partial_i \phi \partial_j \phi +\frac{\delta_{ij}}{2}\partial_k\phi \partial_k\phi +\delta_{ij}\Sigma_0^R\bar{\psi}\gamma^0\psi ,
\eeq
we form the following projections  which will be useful  
\beq
T_{ii}= 3p(r)&=&\frac{1}{4\pi} \sum_{a=1}^A \left[f_a(r)\frac{dg_a(r)}{dr} -\frac{df_a(r)}{dr}g_a(r) -2\kappa_a \frac{f_a(r)g_a(r)}{r}\right] \nn  && +\frac{1}{2}\left(\frac{d\omega}{dr}\right)^2 +\frac{3m^2_\omega}{2}\omega^2  -\frac{1}{2}\left(\frac{d\sigma}{dr}\right)^2 -\frac{3m^2_\sigma}{2}\sigma^2 +\frac{1}{2}\left(\frac{d\rho}{dr}\right)^2 +\frac{3m^2_\rho}{2}\rho^2   +\frac{1}{2}\left(\frac{d\phi}{dr}\right)^2  +3\Sigma_0^R\bar{\psi}\gamma^0\psi, \label{tij2}
\eeq
\beq
s(r)\equiv\frac{3}{2}\left(\frac{r^i r^j}{r^2}-\frac{\delta^{ij}}{3}\right)T_{ij} &=& \frac{1}{4\pi}\sum_{a=1}^A \left[f_a(r)\frac{dg_a(r)}{dr} -\frac{df_a(r)}{dr}g_a(r) + \kappa_a \frac{f_a(r)g_a(r)}{r}\right] \nn
&&  +\left(\frac{d\sigma}{dr}\right)^2
    -\left(\frac{d\omega}{dr}\right)^2
    -\left(\frac{d\rho}{dr}\right)^2
    -\left(\frac{d\phi}{dr}\right)^2. 
    \label{sr}
    \eeq
$f_a,g_a$ are the radial parts of the `large' and `small' components of the nucleon Dirac wave function. Their definitions, together with the derivation of (\ref{sr}), are given in Appendix A. $p(r)$ and $s(r)$ are the `pressure' and   `shear' parts of the energy momentum tensor for spherically symmetric systems  \cite{Polyakov:2002yz}
\beq
T^{ij}=\left(\frac{r^i r^j}{r^2}-\frac{\delta^{ij}}{3}\right)s(r) +\delta^{ij} p(r). 
\label{ps}
\eeq 
From the conservation law $\partial_i T^{ij}=0$, the von Laue condition follows: 
\beq
\int d^3r p(r)= \frac{1}{3}\int d^3r T_{ii}=0.  \label{von}
\eeq

Let us now introduce the D-term for a spin-0 nucleus, defined as the forward limit  of the gravitational form factor $D(t)$
%The latter is defined as, for a spinless nucleus, 
\beq
\langle p'|T^{\mu\nu}|p\rangle = 2A(t) P^\mu P^\nu + \frac{D(t)}{2}(\Delta^\mu\Delta^\nu -g^{\mu\nu}\Delta^2), \label{gra}
\eeq
with $P^\mu= \frac{p^\mu+p'^\mu}{2}$, $\Delta^\mu = p'^\mu -p^\mu$ and $t=\Delta^2$. $A(t=0)=1$ is fixed by momentum conservation, but the value $D=D(t=0)$ is not constrained by any symmetry. 
By the continuity equation, $D$ can be expressed in terms of either the pressure $p(r)$ or the shear $s(r)$ distribution   \cite{Polyakov:2002yz,Polyakov:2018zvc} as
\beq
D=M\int d^3r r^2 p(r) 
=-\frac{4M}{15}\int d^3r r^2 s(r) .
%= -\frac{2M}{5}\int d^3r r^2 \left(\frac{r^i r^j}{r^2}-\frac{\delta^{ij}}{3}\right) T^{ij}.
\label{sint}
\eeq

\section{Nuclear radii and the D-term}

In this section we briefly recapitulate  the various definitions of radii summarized in 
\cite{GarciaMartin-Caro:2023toa} and  to be computed in the next section. 

\begin{enumerate}

\item
Electromagnetic charge radius 

In RMF, protons and neutrons are treated as elementary (pointlike)  Dirac particles. 
In reality, they have their own  electromagnetic form factors reflecting the internal distribution of quarks. The charge radius $r_{em}$ of a nucleus is usually obtained by convoluting the proton and neutron  densities with their respective electromagnetic form factors. Alternatively, one  may first compute the proton number radius 
\beq
\langle r^2\rangle_{p} = \frac{1}{Z}\int d^3r\, r^2 \frac{\rho_B(r)+\rho_{I}(r)}{2}, 
\eeq 
and adopt a simple prescription  \cite{Niksic:2014dra}
\beq
r_{em}=\sqrt{\langle r^2\rangle_p+(0.8)^2}, \label{emp}
\eeq
where the correction reflects the proton size of about 0.8 fm. In \texttt{DIRHB},   (\ref{emp}) reproduces the experimentally measured charge radii at the sub-percent level.
%\MO{(MO comment) Shall we mention here something on the exchange current (two- or multi-body) contribution? for instance, like: Furthermore, it has been pointed out that multi-body contributions (meson-exchange current) are in some cases important (significant).}

\item
Baryon number radius
\beq
\langle r^2\rangle_B = \frac{1}{A} \int d^3r\, r^2 \rho_B(r).  
%= \frac{4\pi}{A}\int_0^\infty dr r^4 \bar{\psi}\gamma^0\psi
\eeq
In contrast to the charge radius,   protons and neutrons equally contribute to $\langle r^2\rangle_B$.  It is therefore a more reasonable measure of the nuclear size, especially for neutron-rich nuclei. In principle, again convolutions with the nucleon form factors are needed. But we neglect them in $\langle r^2\rangle_B$  and in the other radii introduced below. 
\item 
Mass radius
\beq
\langle r^2\rangle_m = \frac{1}{M}\int d^3r\, r^2 T^{00}(r).
%= \frac{4\pi}{M}\int_0^\infty r^4dr T_{00}(r)
\label{mass}
\eeq
While the total mass $M=\int d^3r T^{00}(r)$ is the same for both the canonical and Belinfante-improved energy density, with the weight factor $r^2$ the two energy densities give different results. We will use the Belinfante form which is more appropriate in the context of  GFFs.  
\item 
Scalar  radius
\beq
\langle r^2\rangle_s = \frac{1}{M}\int d^3r\, r^2 T^\mu_\mu(r). \label{sca}
\eeq
In the present model, $T^\mu_\mu$ is given by (\ref{trace}) and represents  the spin-0 part of the energy momentum tensor. 
Using $D$,  we can express the scalar radius as  \cite{Goeke:2007fp}
\beq
\langle r^2\rangle_s=\langle r^2\rangle_m-\frac{3D}{M^2}.
\label{std}
\eeq
\item
Tensor radius 
\beq
\langle r^2\rangle_t = \frac{1}{M}\int d^3r\, r^2 \left(T^{00}(r)+\frac{1}{2}T_{ii}(r)\right) = \frac{3}{2}\langle r^2\rangle_m -\frac{1}{2}\langle r^2\rangle_s.
\label{ten}
\eeq
The linear combination represents the irreducible spin-2 part of the energy momentum tensor \cite{GarciaMartin-Caro:2023toa}. Our definition is different from the one in   \cite{Ji:2021mtz}. Note that $\langle r^2\rangle_t$ is not independent as it is given by \cite{GarciaMartin-Caro:2023toa}
\beq
\langle r^2\rangle_t= \langle r^2\rangle_m+\frac{3D}{2M^2}.
\label{dif2}
\eeq
\item Shear radius 
\beq
\langle r^2\rangle_{shear} = \frac{\int d^3 r\, r^2 s(r)}
{\int d^3 r\, s(r) }=-\frac{15}{4M}\frac{D}{\int d^3r\, s(r)}. \label{shearR}
\eeq
This is analogous to the `mechanical radius' associated with the distribution of the  `normal force' $\frac{2}{3}s(r)+p(r)$ \cite{Polyakov:2018zvc}.  
\end{enumerate}

 We now come to an important issue of the Coulomb field. In fact, if we evaluate $\langle r^2\rangle_m$, $\langle r^2\rangle_t$, $\langle r^2\rangle_{shear}$ and the $D$-term using the formulas derived in the previous section, they diverge due to the Coulomb field. (Notably,  the scalar radius $\langle r^2\rangle_{s}$ is finite.) 
 This is because the Coulomb field of an isolated nucleus decays as %$\phi(r)\cong  \displaystyle\frac{Z\alpha_{em}}{r}$ 
 $\phi(r)\cong  \frac{Ze}{4\pi r}$ 
 at large distances,  and consequently, the integral 
\beq
\int^{r_{max}} d^3r r^2 \left(\frac{d\phi}{dr}\right)^2  \cong Z^2  \alpha_{em} r_{max}, %~~~~~\left( ~~\alpha_{em}=\frac{e^2}{4\pi} ~~ \right)
\eeq
diverges linearly with the cutoff, $r_{max}$. 
In particular, the impact on $D$ is significant
\beq
D_{Coulomb} = \frac{4M}{15} \int \left( \frac{d \phi}{dr}  \right)^2 r^2 d^3r  \cong  \frac{4M}{15}  Z^2 \alpha_{em} r_{max} . \label{dcou}
\eeq 
Despite the suppression by $\alpha_{em}=\frac{e^2}{4\pi}\approx \frac{1}{137}$, (\ref{dcou}) becomes comparable to,  or even dominates over the other contributions to $D$ already when $r_{max}={\cal O}(10)$ fm.  
This is actually a known problem \cite{Donoghue:2001qc,Varma:2020crx}, common to all charged particles (including the proton). Of course, the problem stems from the fact that our starting point contains the QED Lagrangian. Since the GFFs (\ref{gra}) are usually defined through the QCD energy momentum tensor $T^{\mu\nu}=T^{\mu\nu}_{QCD}$, for consistency we ignore the Coulomb field everywhere in the following, including in the equations of motion. A price to pay is that the binding energy and the charge radius deviate  from the successful fit achieved by the full \texttt{DIRHB} calculation. In future, one may introduce a suitable  regulator to effectively include the Coulomb field.

\section{Numerical results}

While the RMF code \texttt{DIRHB} is capable of describing spherical and deformed nuclei alike, in this work we restrict ourselves to spherical even-even nuclei. 
The corresponding RMF code \texttt{DIRHBS} uses a parameter set called  DD-ME2 \cite{Lalazissis:2005de}, summarized in Table \ref{DD-ME2}.
The density-dependent meson-nucleon coupling constants are parameterized as
\beq
&& g_{\sigma} = g_{\sigma}(\rho_B) = g_{\sigma}(\rho_{B,sat}) a_{\sigma} \frac{1+b_{\sigma}(x+d_{\sigma})^2}{1+c_{\sigma}(x+d_{\sigma})^2},~~~~g_{\omega} = g_{\omega}(\rho_B) = g_{\omega}(\rho_{B,sat}) a_{\omega} \frac{1+b_{\omega}(x+d_{\omega})^2}{1+c_{\omega}(x+d_{\omega})^2},  \nonumber  \\
&& g_{\rho} = g_{\rho}(\rho_B) = g_{\rho}(\rho_{B,sat}) \exp \left[ -a_{\rho}(x-1)  \right], \label{grho}
\eeq
where $x = \rho_B(r) / \rho_{B,sat}$ with $\rho_{B,sat} = 0.152~{\rm fm}^{-3}$ being the saturation nuclear density.\footnote{Note that, at $x=1$,  $a_\sigma \frac{1+b_\sigma(1+d_\sigma)^2}{1+c_\sigma(1+d_\sigma)^2}=a_\omega \frac{1+b_\omega(1+d_\omega)^2}{1+c_\omega(1+d_\omega)^2}=1$.}  
The derivatives in (\ref{Sigma0}) are evaluated as, for $k = \sigma, \omega$,
\beq
&& \frac{\partial g_k}{ \partial \rho_B} = \frac{1}{\rho_{sat}}  \frac{\partial g_k}{ \partial x} = \frac{g_{k,sat} a_k}{\rho_{sat}} \left\{  \frac{2b_k (x+d_k)}{1+c_k (x+d_k)^2} - \frac{[1+b_k(x+d_k)^2 ] \cdot 2c_k(x+d_k)}{[1+c_k(x+d_k)^2]^2}  \right\},
\eeq
and similarly for $\partial g_\rho/\partial \rho_B$. 
Together with the meson masses, there are 15 parameters in total  (see Table \ref{DD-ME2}) 
that have been fitted to reproduce nuclear binding energies and charge radii. 
\begin{table}[t]
\begin{tabular}{|c|c|c|c|c|c|c|}
\hline
$m$ [MeV]&$m_{\sigma}$ [MeV]&$m_{\omega}$ [MeV]&$m_{\rho}$ [MeV]
&$g_{\sigma,sat}$&$g_{\omega,sat}$& $g_{\rho,sat}$\\
\hline
$939$&$550.1238$&$783$&$763$&$ 10.5396$&$13.0189$&$3.6836$\\
\hline
\end{tabular}
\vskip 12pt
\begin{tabular}{|c|c|c|c|c|c|c|c|c|}
\hline
$a_{\sigma}$&$b_{\sigma}$&$c_{\sigma}$&$d_{\sigma}$
&$a_{\omega}$&$b_{\omega}$& $c_{\omega}$&$d_{\omega}$&$a_{\rho}$\\
\hline
1.3881&1.0943&1.7057&0.4421&1.3892&0.9240&1.4620&0.4775&0.5647\\
\hline
\end{tabular}
\caption{Values of parameters taken from DD-ME2 \cite{Lalazissis:2005de}. The nucleon mass is fixed.} 
\label{DD-ME2}
\end{table}
With these parameters, \texttt{DIRHBS} solves the Dirac equation (\ref{eom})  and the Poisson equations (\ref{meson}) self-consistently, and builds up nuclei by filling single-particle levels.   Given these solutions, we  compute $T^{00}(r)$, etc., as a function of $r$ and perform one-dimensional integrals over $r$ in the range $0\le r \le 30$ fm. 
We have numerically confirmed that the two integrals (\ref{M}) for the total mass $M$  agree to an accuracy of $0.001\%$. 

%%%%%%%%%%%%%%%%%%%%%%%%%%%%%%%%%%%%%%%%%%%%%%%%%%%%%%%%%%%%%%%%%%%%%%%%%%%%%%%%%%%%%%%%%%%%%
%%%%%%%%%%%%%%%%%%%%%%%%%%%%%%%%%%%%%%%%%%%%%%%%%%%%%%%%%%%%%%%%%%%%%%%%%%%%%%%%%%%%%%%%%%%%%
\begin{table}[h]
\begin{tabular}{|c|c|c|c|c|c|c|c|c|}
\hline
 & $r_{em}$  & $r_B$ &$r_t$ & $r_m$ & $r_s$ & $r_{shear}$ & $D$  &${\cal E}/A$ \\
\hline
$\,^{16}_{\ 8}$O &  2.685 & 2.563 & 2.569 & 2.574  &  2.583 &  2.886 & -90.723  &-8.418 \\
\hline
$\,^{40}_{20}$Ca & 3.387 & 3.291 & 3.295 & 3.303   &  3.318 &  3.766 & -930.453 &-10.408 \\
\hline
$\,^{68}_{28}$Ni  &3.800  &3.862   &3.870  &3.874      &3.881    &4.323 &  -1903.734  &-10.665  \\
\hline
$\,^{90}_{40}$Zr & 4.153  &4.158  &  4.166& 4.170   & 4.178 & 4.586 & -4123.965 &-11.524 \\
\hline
$\,^{132}_{\ 50}$Sn & 4.591  &4.728  & 4.737 &  4.739  & 4.744 & 4.973 & -4192.501 &-11.091 \\
\hline
$\,^{208}_{\ 82}$Pb & 5.323  & 5.448 & 5.454 & 5.459   & 5.468 &  5.926  & -25374.209  &-11.912 \\
\hline
\end{tabular}
\caption{Radii (in units of fm), D-terms (dimensionless) and binding energies per nucleon  (${\cal E}/A$ in units of MeV) of double (sub-)magic nuclei.    $r_{B}\equiv \sqrt{\langle r^2\rangle_{B}}$, etc. $r_{em}$ is calculated from (\ref{emp}). The Coulomb interaction is turned off.
}
\label{tab}
\end{table}
%%%%%%%%%%%%%%%%%%%%%%%%%%%%%%%%%%%%%%%%%%%%%%%%%%%%%%%%%%%%%%%%%%%%%%%%%%%%%%%%%%%%%%%%%%%%%
%%%%%%%%%%%%%%%%%%%%%%%%%%%%%%%%%%%%%%%%%%%%%%%%%%%%%%%%%%%%%%%%%%%%%%%%%%%%%%%%%%%%%%%%%%%%%

\begin{figure} 
\begin{center}
\begin{overpic}[width=0.8\textwidth]{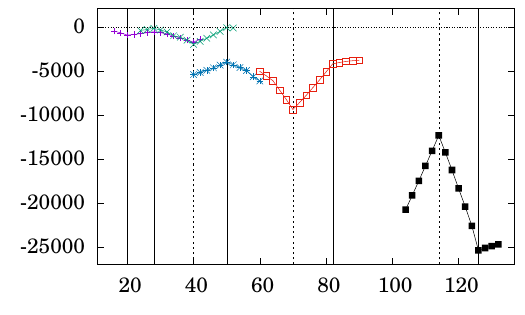}
\put(100,3){\huge $N$}
\put(0,53){\huge $D$}
\put(23.5,50){\color{violet}\large{$\,_{20}$Ca}} 
\put(33,55){\color{teal}\large{$\,_{28}$Ni}} 
\put(36.5,42){\color{cyan}\large{$\,_{40}$Zr}}
\put(56,37){\color{red}\large{$\,_{50}$Sn}}
\put(75,32){\color{black}\large{$\,_{82}$Pb}}
\end{overpic}
\caption{ D-terms of nuclear isotopes as a function of the neutron number $N$. The Coulomb interaction is not included. 
Solid vertical lines indicate the nuclear magic numbers, $N=20,~28,~50,~82,~126$. The dashed lines are at the sub-magic numbers $N=40,~70,~114$.}
\label{DA}
\end{center} \end{figure}
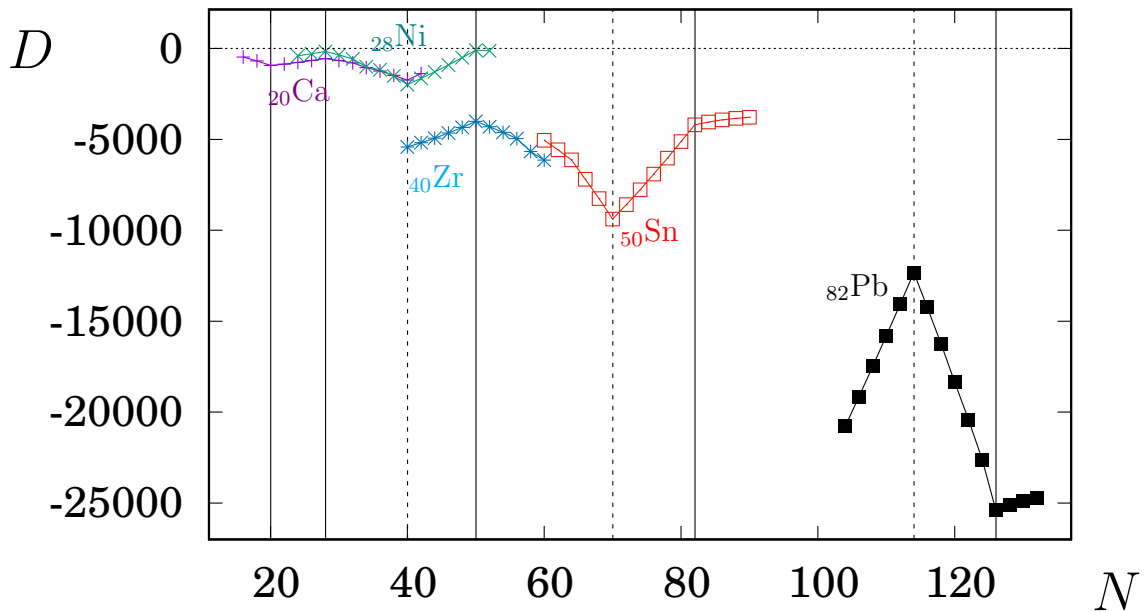

First, we show in Table \ref{tab} the radii and the D-term of double magic nuclei: oxygen (O-16), calcium (Ca-40), zirconium (Zr-90), tin  (Sn-132) and lead  (Pb-208) nuclei.   Simpler notations $r_{B}\equiv \sqrt{\langle r^2\rangle_{B}}$, etc., are used here and in the following.  $D$ is computed from $p(r)$ in (\ref{sint}).\footnote{The values of $D$ computed from the two integrals in (\ref{sint}) differ by a few percent for most nuclei studied in this work. This can be interpreted as an estimate of uncertainty  in the present calculation. We however note that the difference becomes larger ($> 10\%$) for certain isotopes of Ni that have unusually small values of $|D|$, see Fig.~\ref{DA}. We think that $p(r)$ is more reliably computed in the current \texttt{DIRHB} code  because the integral (\ref{von}) vanishes with  excellent accuracy ${\cal O}(10^{-5})$ in all cases.   }   Since the Coulomb field is neglected, the charge radii are smaller than the experimental values by a few percent. Also, it is known that, without the Coulomb interaction, the binding energy per nucleon is significantly larger in magnitude, especially for larger nuclei. (For example, the experimental value for $\,^{208}$Pb is ${\cal E}/A=-7.87$ MeV.) After  including the Coulomb field, \texttt{DIRHB} achieves  percent-level precision for these observables. 

As for the other radii, we find the ordering 
\beq
r_B< r_t < r_m < r_s<r_{shear}, 
\eeq
for all the nuclei in Table~\ref{tab}.  The five radii $r_{em,B,t,m,s}$ are within a few percent to each other, whereas  $r_{shear}$ is larger by about 10\%.  
The ordering $r_t<r_m<r_s$ is a direct consequence of the negativity of $D$, whose magnitude grows substantially with increasing $A=Z+N$. Such a rapid increase  has been previously observed \cite{Polyakov:2002yz,Guzey:2005ba,Liuti:2005qj,GarciaMartin-Caro:2023klo,GarciaMartin-Caro:2023toa}.
Our results are smaller in magnitude than those obtained in the Walecka model \cite{Guzey:2005ba},  where a direct comparison is possible.  (Note that our $D$ is related to $d_A$ in \cite{Guzey:2005ba} as $D=\frac{4}{5}d_A$.)  For $N\neq Z$ nuclei, this can be mainly attributed to the inclusion of the $\rho$-meson which contributes positively to the D-term.    In the literature, the   $A$-dependence is often parametrized as  
\beq
D\propto A^\beta.  \label{naive}
\eeq
In the liquid drop model, there is an analytical result $\beta = \frac{7}{3}$  
\cite{Polyakov:2002yz}. In  the Walecka model,  $\beta\approx 2.26$ \cite{Guzey:2005ba}, whereas  in the Skyrme model,   $\beta\approx 1.9$ \cite{GarciaMartin-Caro:2023klo,GarciaMartin-Caro:2023toa}. However, Table~\ref{tab} already suggests a tension with such a parametrization. The D-terms of $\,^{90}$Zr and $\,^{132}$Sn differ by less than 2\%, although their mass numbers $A$ differ by almost 50\%. One of the main findings of the present work is that the simple formula (\ref{naive}) does not capture the strong sensitivity of the D-term to the nuclear shell structure. To demonstrate this, in Fig.~\ref{DA}, we plot the D-terms of the Ca, Ni, Zr, Sn, Pb isotopes as a function of the neutron number $N$. We see that $|D|$ does not monotonically increase with $A$. Instead, there are local maxima and minima when the neutron number coincides with the magic numbers $N=20,~28,~50,~82,~126$ (solid vertical lines), and even at the `sub-magic' numbers  $N=40,~70,~114$  (dashed vertical lines)  associated with the closure of the $1p_{\frac{1}{2}}$,  $1d_{\frac{3}{2}}$ and $0i_{13/2}$ ($l=6$) levels,  respectively.\footnote{ Note that these sub-shell structures depend on the interaction. The present result pertains to the DD-ME2 set.} If one associates the negativity of the D-term with stability, one might naively  expect the D-term to exhibit a local maximum (in magnitude) at magic numbers.  However, the most stable Zr nucleus, $^{90}_{40}$Zr, has the smallest $|D|$ among its  isotopes. Moreover, the nickel isotope with the most negative D-term,  $\,^{68}_{28}$Ni, is unstable against weak decay, despite being doubly (sub-)magic. An apparent rule  one can deduce from Fig.~\ref{DA} is that  maxima and minima  alternate from one (sub-)magic number to another. Such a behavior has to do with the fact that the D-term is sensitive to the properties of individual nucleon orbits $f_a,g_a$, see (\ref{tij2}). We leave its microscopic explanation  for future work.

%%%%%%%%%%%%%%%%%%%%%%%%%%%%%%%%%%%%%%%%%%%%%%%%%%%%%%%%%%%%%%%%%%%%%%

%%%%%%%%%%%%%%%%%%%%%%%%%%%%%%%%%%%%%%%%%%%%%%%%%%%%%%%%%%%%%%%%%%%%%%
\begin{figure}[t] \begin{center}
\begin{overpic}[width=0.7\textwidth]{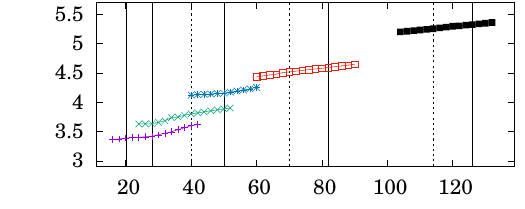}
\put( 3,20){\rotatebox{90}{\Large $r_{em}$ [fm]}}
\put(19,15){\color{violet}\large{$\,_{20}$Ca}} 
\put(29,19){\color{teal}\large{$\,_{28}$Ni}} 
\put(36,24){\color{cyan}\large{$\,_{40}$Zr}}
\put(50,29){\color{red}\large{$\,_{50}$Sn}}
\put(76,30){\color{black}\large{$\,_{82}$Pb}}
\put(100, 3){\Large $N$}
\end{overpic}
\begin{overpic}[width=0.7\textwidth]{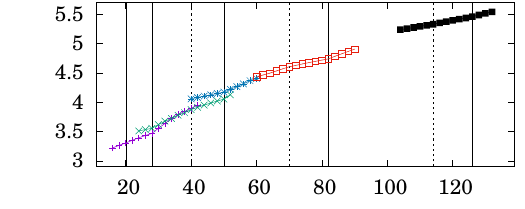}
\put(18,15){\color{violet}\large{$\,_{20}$Ca}} 
\put(27,19){\color{teal}\large{$\,_{28}$Ni}} 
\put(36,24){\color{cyan}\large{$\,_{40}$Zr}}
\put(50,30){\color{red}\large{$\,_{50}$Sn}}
\put(79,31){\color{black}\large{$\,_{82}$Pb}}
\put( 3,20){\rotatebox{90}{\Large $r_{m}$ [fm]}}
\put(100, 3){\Large $N$}
\end{overpic}
\begin{overpic}[width=0.7\textwidth]{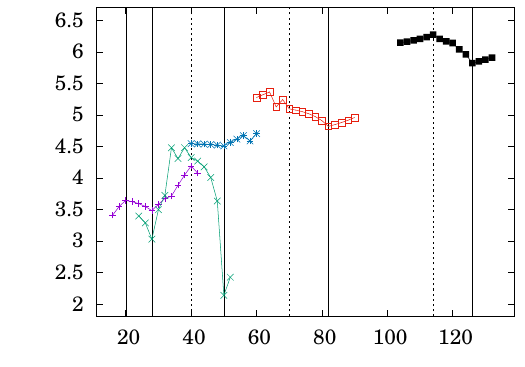}
\put( 3,31){\rotatebox{90}{\Large $r_{shear}$ [fm]}}
\put(20,33){\color{violet}\large{$\,_{20}$Ca}} 
\put(33,29){\color{teal}\large{$\,_{28}$Ni}} 
\put(40,46){\color{cyan}\large{$\,_{40}$Zr}}
\put(55,41){\color{red}\large{$\,_{50}$Sn}}
\put(77,56){\color{black}\large{$\,_{82}$Pb}}
\put(100, 4){\Large $N$}
\end{overpic}
\caption{Charge radius (top), mass radius (middle) and shear  radius (bottom) of Ca, Ni, Zr, Sn and Pb isotopes.}
\label{rmplot}
\end{center} \end{figure}
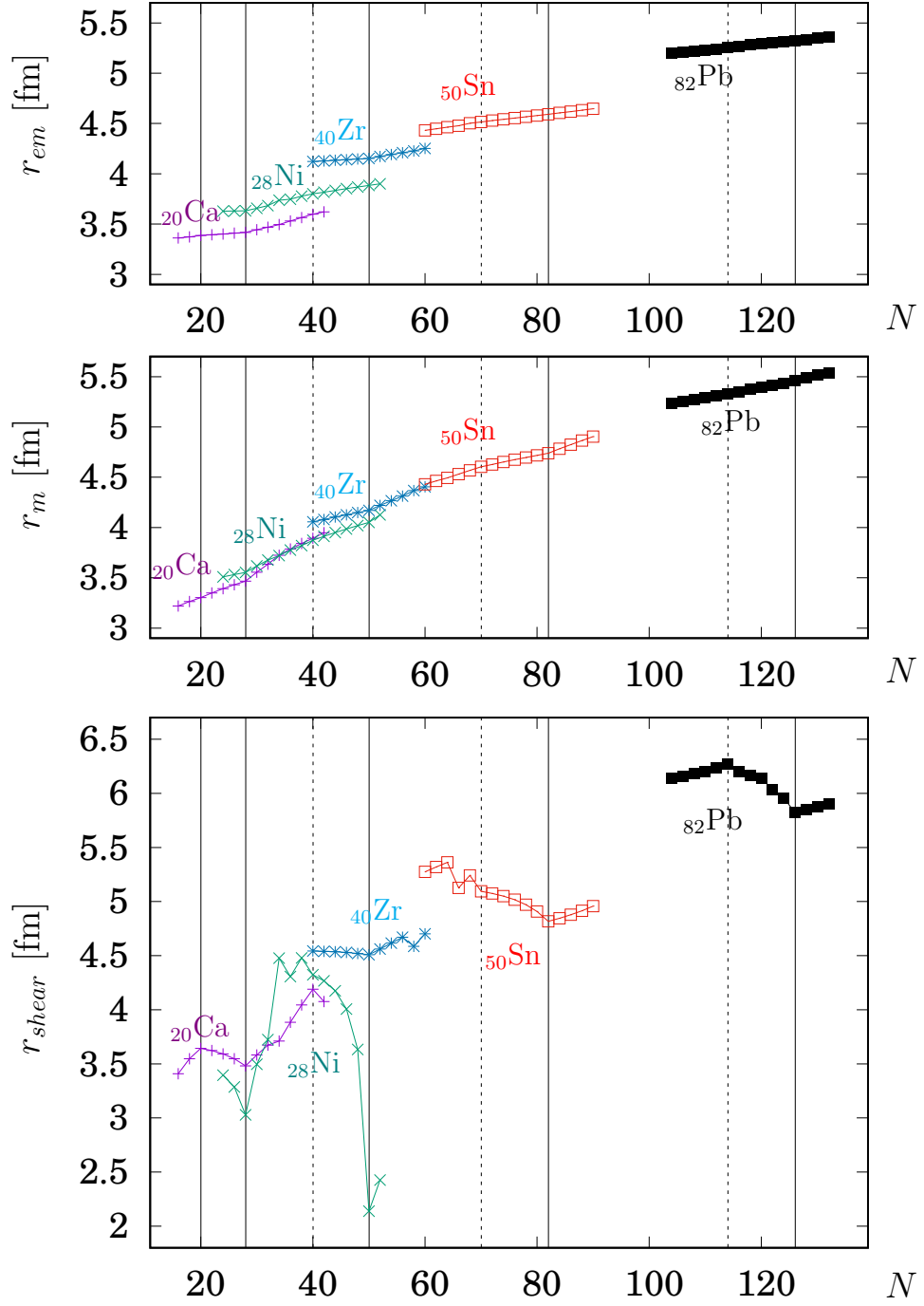
%%%%%%%%%%%%%%%%%%%%%%%%%%%%%%%%%%%%%%%%%%%%%%%%%%%%%%%%%%%%%%%%%%%%%%

Next, in Fig.~\ref{rmplot}, we plot the charge radii $r_{em}$, the mass radii $r_m$,  and the shear radii $r_{shear}$ of the same  isotopes as a function of the neutron number $N$.
%In Fig.~\ref{rescale}, we plot the rescaled mass radius $r_m/A^{1/3}$, together with  the charge and baryon number counterparts. 
We recognize conspicuous shell effects manifesting here as kinks at magic numbers, and to a lesser extent, also at sub-magic numbers. Such kinks in isotope shifts are well known for the charge radius 
(see, e.g., \cite{Perera2021ChargeRadii,Naito:2022vnz,Konig:2023bzp}), but this is the first observation of a similar structure in the mass radius. In the shear radius, the effect is so strong that the radius even  {\it decreases} with increasing $N$ in certain intervals of $N$. This is  correlated with the zigzag behavior of $D$ as can be seen from (\ref{shearR}), but the integral $\int d^3r s(r)$ in the denominator also varies strongly with $N$.   
The distinct  outlier in the nickel isotope chain, $^{78}_{28}$Ni, is an interesting nucleus in its own    \cite{Taniuchi:2019pen,Hagen2019DoublyMagic}. It is one of the heaviest known doubly magic nuclei far from stability. The unusually small shear radius $r_{shear}\approx 2.1$ fm and the  D-term $|D|\approx 90$ further highlight the intriguing nature of this nucleus.

It is also interesting to notice that the charge radius grows slower than $A^{1/3}$, obviously because adding neutrons does not directly expand  the distribution of charges. However, the mass radius keeps up with the $A^{1/3}$ scaling because protons and neutrons equally contribute to the mass distribution. The shear radius grows stronger than $A^{1/3}$ on average. A similar observation was made for the mechanical radius in the Skyrme model \cite{GarciaMartin-Caro:2023toa}. 

%%%%%%%%%%%%%%%%%%%%%%%%%%%%%%%%%%%%%%%%%%%%%%%%%%%%%%%%%%%%%%%%%%%% 
\begin{figure}
\begin{overpic}[width=0.6\textwidth]{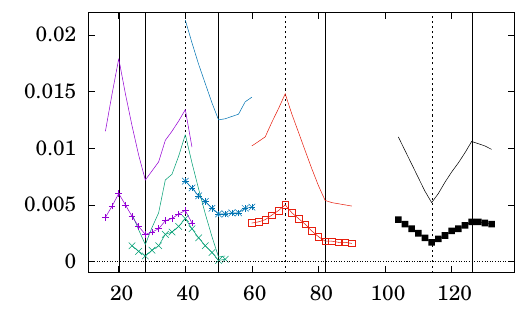}
\put(18,25){\color{violet}\large{$\,_{20}$Ca}} 
\put(30,11){\color{teal}\large{$\,_{28}$Ni}} 
\put(38,29){\color{cyan}\large{$\,_{40}$Zr}}
\put(50,25){\color{red}\large{$\,_{50}$Sn}}
\put(82,19){\color{black}\large{$\,_{82}$Pb}}
         \put(102,3){\Large  $N$}
    \put(40,61){\Large{$r_s-r_t$} and \Large{$r_m-r_t$} [fm]}
\end{overpic} 
\caption{Differences $r_s-r_t$ (upper lines) and $r_m-r_t$ (lower lines with symbols) as a function of the neutron number $N$.}
\label{diff}
\end{figure}
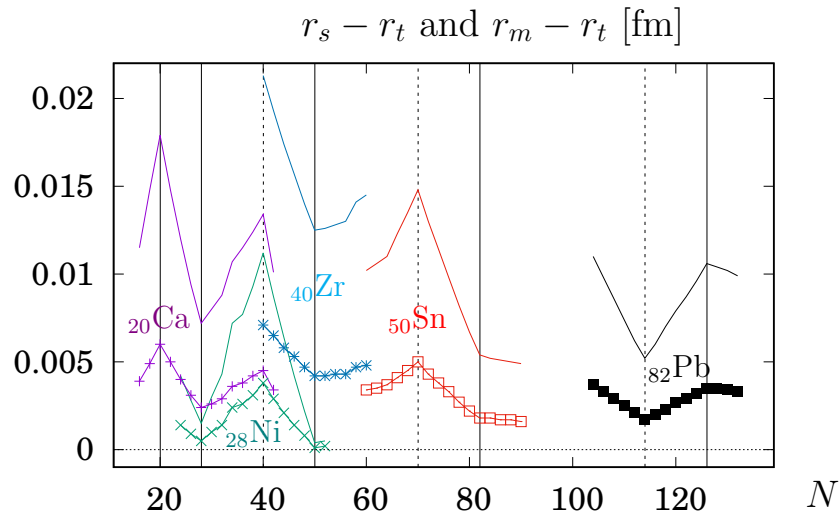
%%%%%%%%%%%%%%%%%%%%%%%%%%%%%%%%%%%%%%%%%%%%%%%%%%%%%%%%%%%%%%%%%%%% 

Since the scalar radius $r_s$ and tensor radius $r_t$ are numerically very close to $r_m$,  we plot the differences $r_s-r_t$ and $r_m-r_t$ in Fig.~\ref{diff}. 
From (\ref{std}) and (\ref{dif2}), we find that  
\beq
r_s-r_t,\  r_m-r_t \sim {\cal O}\left(\frac{D}{M^2 r_m}\right),
\eeq
Evidently, the strong shell effect in the D-term, Fig.~\ref{DA}, is directly reflected. 
When averaged over  isotopes, the differences are  roughly constant in $N$. This suggests that $D\propto  M^2r_m \propto A^{\beta=7/3}$ on average, and in this sense our result is consistent with those in the liquid drop model  \cite{Polyakov:2002yz} and the Walecka model \cite{Guzey:2005ba}.   However, because of the strong oscillation, it seems pointless to try to determine $\beta$ more accurately than $\beta \sim 2$. The Skyrme model result $\beta=1.9$ \cite{GarciaMartin-Caro:2023klo,GarciaMartin-Caro:2023toa} gives an equally good fit.

\begin{figure}[t] \begin{center}
\begin{overpic}[width=0.45\textwidth]{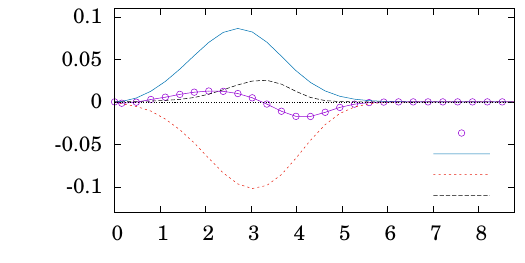}
\put(3,16){\rotatebox{90}{$r^2p(r)$ [GeV/fm]}}
%%%\put(50,1){\large{$r\,$ [fm]}}
\put(80,40){\Large{$\,^{40}$Ca}}
\put(65,24){total}
\put(65,20){nucleons}
\put(65,16){mesons}
\put(65,12){rearrange}
\end{overpic}
\hspace{1mm}
\begin{overpic}[width=0.45\textwidth]{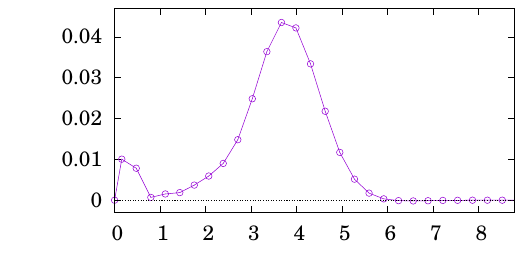}
\put(3,16){\rotatebox{90}{$r^2s(r)$ [GeV/fm]}}
\put(80,40){\Large{$\,^{40}$Ca}}
%%%\put(50,1){\large{$r\,$ [fm]}}
\end{overpic}
\begin{overpic}
[width=0.45\textwidth]{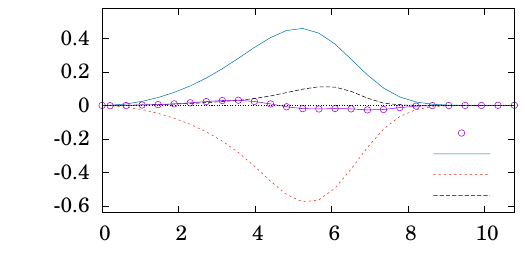}
\put(3,16){\rotatebox{90}{$r^2p(r)$ [GeV/fm]}}
\put(50,-1){$r\,$ [fm]}
\put(80,40){\Large{$\,^{208}$Pb}}
\put(65,24){total}
\put(65,20){nucleons}
\put(65,16){mesons}
\put(65,12){rearrange}
\end{overpic}
\begin{overpic}
[width=0.45\textwidth]{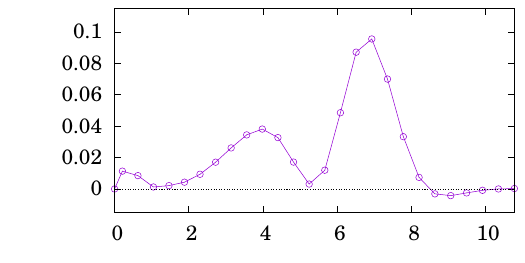}
\put(3,16){\rotatebox{90}{$r^2s(r)$ [GeV/fm]}}
\put(55,-1){$r\,$ [fm]}
\put(80,40){\Large{$\,^{208}$Pb}}
\end{overpic}
\caption{Top: Pressure and shear distributions, $r^2 p(r)$ (left) and $r^2s(r)$ (right), of $^{40}$Ca. Bottom: the same for $^{208}$Pb. 
The plots of $r^2p(r)$ also show the individual  contributions  from the nucleon, mesons, and rearrangement terms, see (\ref{tij2}).}
\label{psplot}
\end{center} \end{figure}
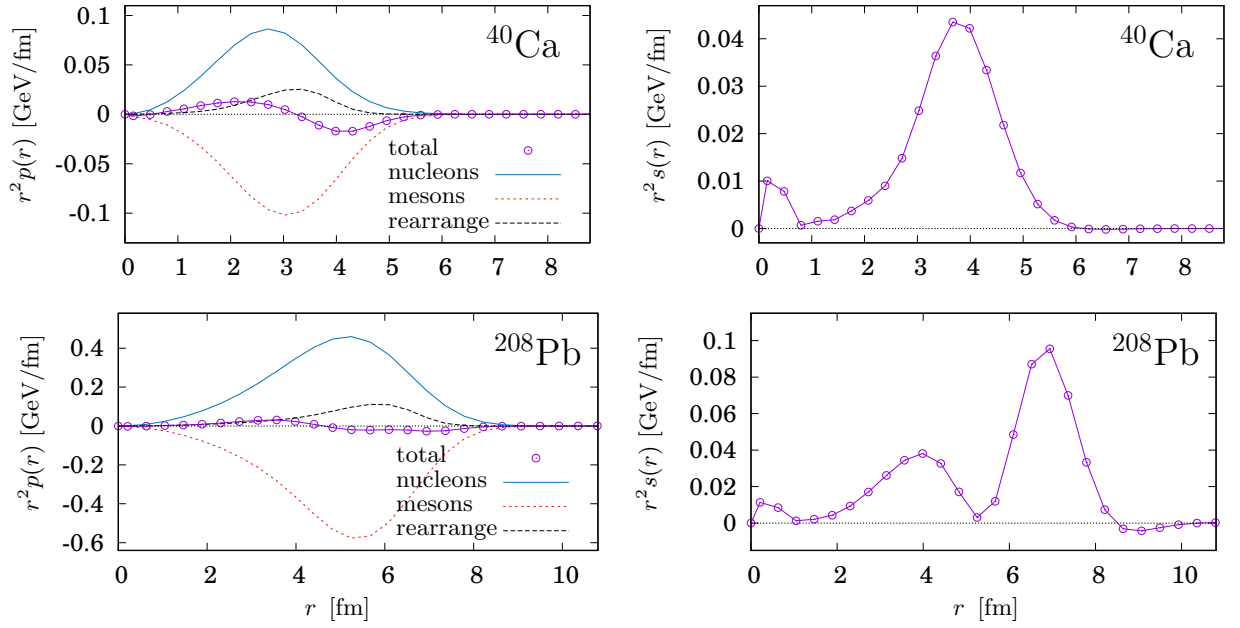

Finally, in Fig.~\ref{psplot}, we plot the pressure $p(r)$ and shear $s(r)$ distributions   of $^{40}$Ca (top) and $^{208}$Pb (bottom). Nucleons  contribute positively to $p(r)$, while the total meson contribution is negative due to the strong attraction from the $\sigma$-meson. 
We have numerically  confirmed that the von Laue condition (\ref{von}) is satisfied to a good accuracy: The   integral $\int d^3r p(r)$ is about 0.01\% of the total nuclear mass $M$. 
A large  cancellation between the  nucleon  and meson  contributions is the dominant mechanism to achieve this mechanical equilibrium condition,  
but  the rearrangement term $\sim \Sigma_0^R\bar{\psi}\gamma^0\psi$ (see the last term in  (\ref{tij2}))  is also numerically important.  In contrast, the shear distribution is almost everywhere positive. The peak position of $r^2s(r)$ roughly coincides with $r_{shear}$. 

%Previously, the Skyrme model calculation found $p(r\sim 0)<0$ in the core region for all nuclei $A>1$ \cite{GarciaMartin-Caro:2023toa}, and such a negative behavior was also found in specific models for  the deuteron and the helium-4 nucleus \cite{Freese:2022yur,He:2023ogg,He:2024vzz}. 

%%%%%%%%%%%%%%%%%%%%%%%%%%%%%%%%%%%%%%%%%%%%%%%%%%%%%%%%%%%%%%%%%%%%%%

%%%%%%%%%

%%%%%%%%%%%%%%%%%%%%%%%%%%%%%%%%%%%%%%%%%%%%%%%%%%%%%%%%%%%%%%%%%%%%%%

\section{Conclusions} 

In this paper,  we have used the RMF theory of atomic nuclei to study the mechanical structure of medium-sized or larger nuclei. The previous work  \cite{Guzey:2005ba} computed the D-term of several nuclei in the Walecka model. We have used a more sophisticated version of RMF and systematically computed the D-term and the various definitions of radius across a large number of nuclear isotopes.  Given that the \texttt{DIRHB} framework successfully describes the ground state properties of nuclei to within a few percent, we can expect that our predictions  are accurate at a similar level of  precision. In particular, we have observed, for the first time, the strong oscillation of the D-term of nuclear isotopes as the neutron number is varied. The sharp kinks at the magic numbers and even sub-magic numbers indicate that the D-term and the associated radii are sensitive to the nuclear shell structure and individual nucleon orbits.

There are a number of directions left for future studies. In this work we only considered spherical even-even nuclei. RMF packages for deformed nuclei, \texttt{DIRHBZ} and \texttt{DIRHBT}, are available \cite{Niksic:2014dra}, but additional care is  needed for the extraction of radii and form factors in the presence of deformation, cf.,  \cite{GarciaMartin-Caro:2023klo}. In principle, these codes can also be used for even-odd or odd-odd nuclei, but this requires further technical and numerical improvements. 
Another important effect  is the pairing correlation.   While the \texttt{DIRHB} codes by default feature the relativistic Hartree-Bogoliubov (\texttt{HB}) framework, this has been turned off in our calculation. For non-magic nuclei, the pairing effect fine-tunes the binding energy at the percent level.  Its impact on the mechanical properties of nuclei has not been explored in the literature.

Finally, it remains an open question  how  the various definitions of radii calculated in this work can be measured experimentally. Ideally, such  measurements should be accurate enough to discriminate the percent-level differences between the different definitions of radius that we have predicted. This is however  extremely challenging due to the lack of a clean probe (`graviton') of the distributions of energy and momentum  inside nuclei. There have been a few recent proposals for observables 
\cite{Hatta:2023fqc,Hagiwara:2024wqz,He:2024lry}, but more efforts in this direction are  certainly necessary. 

\section*{Acknowledgments}
We are grateful to Masaaki Kimura for collaboration in the early stage of this work. 
We also thank Tamara Nik\v{s}i\'c, Peter Schweitzer and Marat Siddikov for discussion and  correspondence. Y.~H. was supported by the U.S. Department
of Energy under Contract No. DE-SC0012704, and also by LDRD funds from Brookhaven Science Associates.  
M.~O is supported in part by JSPS Grants-in-Aid for Transformative Research Areas 
(Quantum Matter Science in the Universe Opened Up by Precise Numerical Calculations), 
JP25A203 and JP25H01267, and also by
JSPS Grants-in-Aid for Scientific Research JP23K03427.

\appendix 

\section{Solution of the Dirac equation in a spherical potential}

We work with the standard representation of the gamma matrices 
\beq
\gamma^0=\begin{pmatrix} 1 & 0 \\ 0 & -1 \end{pmatrix} ,\qquad \gamma^i = \begin{pmatrix} 0 & \sigma^i \\ -\sigma^i & 0 \end{pmatrix} = -\gamma_i,
\eeq
 and parametrize the solution of the Dirac equation (\ref{eom}) in the form  
\beq
\psi_a = \begin{pmatrix} f_a(r)\Phi^\pm_{jm} \\ ig_a(r) \Phi^\mp_{jm}\end{pmatrix} . %\qquad \bar{\psi}_a= \left(f_a(r)\Phi^{\pm\dagger}_{jm}, i g_a(r)\Phi^{\mp \dagger}_{jm}\right).
\eeq
$\Phi^\pm_{jm}$ is the spinor spherical harmonics with orbital angular momentum $l=j\mp \frac{1}{2}$ defined by 
\beq
\Phi^\pm_{jm}(\theta,\varphi)= \sum_{m_s, m_{l}} \langle\, \frac{1}{2},m_s;l=j\mp \frac{1}{2}, m_{l}\,|\,jm\rangle
\,\chi_{m_s} Y_{l m_{l}}(\theta,\varphi). \label{cle}
\eeq
The two-component  spinors $\chi_{m_s}$  and the spherical harmonics $Y_{l m}$ are combined via the Clebsch-Gordan coefficients. 
The phase of the wave function is chosen so that $f_a$ and $g_a$ are real and they are normalized as  $1=\int_0^\infty dr r^2(f_a^2(r)+g_a^2(r))$. 
The parameter $\kappa$ is assigned for  each shell orbit as  
\beq
\kappa_a = \begin{cases}  -(j+\frac{1}{2}) \qquad l=j-\frac{1}{2} \\ 
j+\frac{1}{2} \qquad l=j+\frac{1}{2} \end{cases}
\eeq

To derive (\ref{tij2}) and  (\ref{sr}), the following formulas are in order
\beq
\vec{\sigma}\cdot \hat{r}\Phi^\pm_{jm} = + \Phi^\mp_{jm} \qquad (\vec{\sigma}\cdot \vec{L}+1) \Phi^\pm_{jm} = \pm (j+\frac{1}{2})\Phi^\pm_{jm} \qquad 
\vec{\sigma}\cdot \vec{\nabla} = \vec{\sigma}\cdot \hat{r} \left(\frac{\partial}{\partial r} - \frac{1}{r} \vec{\sigma}\cdot \vec{L}\right)
\eeq
Note that the sign of the first formula is specific to the present convention (\ref{cle}).  (The ordering of $\frac{1}{2}$ and $l$ matters.) 
We find 
\beq
i \,\delta_{ij}\,\gamma_i\partial_j\psi
%=-i \,g^{ij}\,\gamma_i\partial_j\psi
=
\begin{pmatrix}  \vec{\sigma}\cdot \vec{\nabla} g\Phi^\mp_{jm}  \\ i \vec{\sigma}\cdot \vec{\nabla} f\Phi^\pm_{jm} \end{pmatrix}  =  \begin{pmatrix} 
\left(g' + \frac{g}{r}(1\pm (j+\frac{1}{2})) \right)\Phi^\pm_{jm} \\ i\left(f'+\frac{f}{r}(1\mp (j+\frac{1}{2}))\right) \Phi^\mp_{jm}  \end{pmatrix}
\eeq
such that 
\beq
\delta_{ij}\,\frac{i}{2}\bar{\psi}\gamma_i (\partial_j-\overleftarrow{\partial}_j)\psi &=& f \left(g' + \frac{g}{r}(1\pm (j+\frac{1}{2}))\right)|\Phi^\pm|^2 - g \left(f'+\frac{f}{r}(1\mp (j+\frac{1}{2}))\right) |\Phi^{\mp}|^2 \nn 
&\to& \frac{1}{4\pi} \left(fg'-f'g\pm \frac{2fg}{r}(j+\frac{1}{2})\right). 
\eeq
We have effectively replaced  $|\Phi|^2\to \frac{1}{4\pi}$, anticipating angular integration for spherical nuclei.  
Similarly, 
\beq
\frac{r^ir^j}{r^2} \frac{i}{2}\bar{\psi}\gamma_i(\partial_j-\overleftarrow{\partial}_j)\psi = \frac{i}{2r}r^i\bar{\psi}\gamma_i (\partial_r - \overleftarrow{\partial_r})\psi &=& \frac{r^i}{2r}(fg'-f'g) \left(\Phi^{\pm*}_{jm} \sigma^i \Phi^\mp_{jm} + \Phi^{\mp*}_{jm} \sigma^i \Phi^\pm_{jm}\right) \nn 
&=&\frac{1}{2}(fg'-f'g)\left(| \Phi^\mp_{jm}|^2 + | \Phi^\pm_{jm}|^2\right) \nn 
&\to & \frac{fg'-f'g}{4\pi}\label{y}
\eeq

\bibliography{ref}

\end{document}